\documentclass[sn-mathphys]{sn-jnl}

\jyear{2021}%

\theoremstyle{thmstyleone}%

\theoremstyle{thmstyletwo}%
\usepackage{physics}
\theoremstyle{thmstylethree}%
\usepackage{amsmath}
\usepackage{bbold}
\usepackage{tabularx}
\begin{document}

\title[Combating errors in propagation of orbital angular momentum $\cdots$]{Combating errors in propagation of orbital angular momentum modes of light in turbulent media}

\author*[]{\fnm{Rajni} \sur{Bala}}\email{Rajni.Bala@physics.iitd.ac.in}

\author[]{\fnm{Sooryansh} \sur{Asthana}}\email{sooryansh.asthana@physics.iitd.ac.in}

\author[]{\fnm{V.} \sur{Ravishankar}}\email{vravi@physics.iitd.ac.in}

\affil*[]{\orgdiv{Department of Physics}, \orgname{Indian Institute of Technology Delhi}, \orgaddress{\street{Hauz Khas}, \city{New Delhi}, \postcode{110016}, \country{India}}}
\abstract{ There is a wealth of  simulation, experimental and analytical  studies on propagation of orbital angular momentum (OAM) modes through atmospheric and oceanic turbulence.} Using the data of these studies and generalising  the framework proposed in  [Bala {\it et al.}, \href{arXiv:2208.04555}{https://arxiv.org/abs/2208.04555}] for error-immune information transfer, we accomplish two tasks. First, we identify invariants for propagation of OAM modes through atmospheric and oceanic turbulence, in which error-immune information can be encoded. A closer look at the data reveals two universal features: (i)  coherence lasts for a much longer distance in turbulence than entanglement, and, (ii) the crosstalk among different OAM modes depends very weakly on the initial OAM mode index in the weak turbulence regime.  Keeping these in mind, we next develop a method for combating errors in what we call an idealised crosstalk channel. In an  idealised crosstalk channel, the  crossover probabilities are independent of the initial mode index (IMI). We lay down a procedure that allows to retrieve full information in a state by identifying invariant quantities.  Finally, we construct quantum error correction and rejection codes for idealised crosstalk channels, without any need for multiparty entanglement.


\maketitle
\section{Introduction}
 \label{Introduction}  
  Recent times have witnessed an unprecedented surge of interest in quantum information-theoretic protocols \cite{Bennett84, Ekert91, Bennett93, Bennett2001remote}. 
  However, noise acts as a major impediment in their implementation. Though there have been many proposals \cite{steane1996simple, schumacher2002approximate, zanardi1997error, layden2019ancilla, lidar1998decoherence, viola1999dynamical} for mitigating the effect of noise, these techniques either demand resources such as multiparty entanglement or  implementation of interventions  at appropriate time intervals \cite{byrd2004overview}. Generation of multiparty entanglement is still a challenge, especially in  photonic systems \cite{proietti2021experimental}.

   These challenges in implementation of conventional error mitigation techniques have led us to propose a new information encoding scheme \cite{Bala22}. This scheme rests on encoding information in invariants, due to which information can be transferred in an error-immune manner even through noisy channels.

 In parallel, thanks to significant advancements in generation and measurement of OAM states of light, these modes hold the promise of implementation of communication protocols with flying qudits, for both-- free-space and underwater  communication \cite{mirhosseini2015high,wang2020satellite,li2020deterministic, erhard2020advances}.  However, turbulence, being atmospheric or oceanic, acts as a major impediment \cite{bachmann2019universal, zhai2020effects} leading to crossover among different modes and decoherence/ loss of entanglement. The propagation of OAM modes through turbulence is naturally modelled through cross-talk channels-- which cause a spillover to neighbouring modes. That is to say, it results in an increase in number of modes in the final state  as compared to the initial state.

In this work, we apply the framework proposed in \cite{Bala22} to address the problem of transmission of information with OAM modes through turbulent media.  There are two physical situations of interest, {\it viz.}, atmospheric and oceanic turbulence. To start with, we analyse data from simulation,  experimental and analytical studies on propagation of  OAM modes of light \cite{anguita2008turbulence, Bachmann_2019, leonhard2015universal, zhai2020effects, yan2017effect}. The data serve a twofold purpose: (i) identification of invariants for transmission of error-free information with OAM modes through turbulent media, (ii) a closer look at the data reveals  a  weak dependence of crosstalk on the initial mode index (IMI) in weak turbulence regime. Current simulations have studied propagation of a pure entangled  biphoton and three-photon OAM state which mimic `two-qubit' and `three-qubit' respectively. We study  information retrieval for both the  cases (section (\ref{relation})).
Interestingly,  the invariants identified in the case of atmospheric turbulence also allow for retrieval of full information in the initial state.

 We briefly summarise the results of simulation,  experimental and analytical studies. It is found  \cite{paterson2005atmospheric, leonhard2015universal, tyler2009influence} that  in weak atmospheric turbulence,  the crossover probabilities are independent of the IMI under rather simplified assumptions. In a more realistic simulation study, it is observed \cite{anguita2008turbulence} that for weak turbulence $(C_n^2\approx 10^{-16} {\rm m}^{-2/3})$, crosstalk is negligible for all practical purposes. For $C_n^2 =10^{-15}{\rm m}^{-2/3},$ the crosstalk is independent of the IMI for $2\leq \ell \leq 7$.  By examining the data presented in \cite{anguita2008turbulence}, we  find out  that the crosstalk among different modes depends only weakly  on the IMI in another regime characterised by $6 \leq \ell \leq 16$ and $C_n^2 = 10^{-14} m^{-2/3}$ (see section (\ref{Weak_turbulence}) for details).

 It is observed in \cite{zhai2020effects} for propagation of `two-qubit' OAM Werner state that  coherence lasts much longer than   entanglement (the typical distances are of the orders of $50$ $m$ and $10$ $m$ respectively). Similar conclusions have been drawn in \cite{wei2017universal} for propagation of `two-qubit' OAM Werner state through atmospheric turbulence.
 
 From these findings, we may conclude that, in weak turbulence regime, the noisy channels are mostly {\it ideal}, by which we mean that the crossover probabilities are independent of the IMI. Exploiting this feature, we  identify invariants for an IC channel (section(\ref{Framework})). We  also demonstrate that full information retrieval is possible by employing these invariants (section(\ref{information})).
 
 Finally, keeping in mind the experimental challenges in generation of multiparty entanglement among OAM modes and its fast degradation through turbulence, in this work, we also construct {\it ancilla-free} error-rejection and error-correction codes with appropriate initial modes (resp. sections (\ref{rejection}) and (\ref{correction})) for an IC channel. Section (\ref{conclusion}) concludes the paper with closing remarks.
\section{Propagation of OAM modes in crosstalk channels: retrieval of information}
\label{relation}
There are a number of studies which give final density matrix (in a truncated subspace) after propagation through atmospheric or oceanic turbulence resulting in crossover, decoherence and loss of entanglement. In spite of this, we show that some information can be retrieved in terms of quantities that remain invariant. For this, we consider various simulation and analytical studies of transfer of information in OAM modes   through weak turbulence  \cite{ leonhard2015universal, zhai2020effects, yan2017effect}. Employing the results of these studies, we identify invariants which, in some of the cases, lead to retrieval of full information contained in the initial state. 
To start with, we consider  propagation of Werner state propagating through oceanic turbulence. 

\subsection{Retrieval of information from a Werner-like state after passing through oceanic turbulence}
 In \cite{zhai2020effects}, Zhai. {\it et al.} have studied the effect of oceanic non-Kolmogorov turbulence on propagation of entangled Laguerre-Gauss beams by chosing the initial state of the photon pair to be in a Werner-like state,
\begin{align}
    \rho^{(0)}=\Big(\frac{1-\gamma}{4}\Big)\mathbb{1}+\gamma\vert\psi_0\rangle\langle \psi_0\vert.
    \label{Initial_rho0}
\end{align}
The symbol $ \gamma$ characterizes the purity of the initial state. The symbol  $\mathbb{1}$ represents the identity matrix in the subspace spanned by $\{\ket{\ell,\ell}, \ket{\ell,-\ell}, \ket{-\ell, \ell}, \ket{-\ell, -\ell}\}$  and the
state $\vert\psi_0\rangle$ is chosen to be,
\begin{align}
\vert\psi_0\rangle =\cos\frac{\theta}{2}\vert\ell,-\ell\rangle+e^{i\phi}\sin\frac{\theta}{2} \vert -\ell,\ell\rangle.
\end{align}

In \cite{zhai2020effects},  the same turbulence effect has been assumed on both the photons for short distances. They determine the final state in the ordered  trucated basis $\{\ket{\ell,\ell},\ket{\ell,-\ell},\ket{-\ell,\ell},\ket{-\ell,-\ell}\}$ and report that  the non-vanishing elements are given by,
     \begin{eqnarray}         &&\rho_{11}=\alpha\Big(\mu^2\rho_{11}^{(0)}+\mu\nu\rho_{22}^{(0)}+\mu\nu\rho_{33}^{(0)}+\nu^2\rho_{44}^{(0)}\Big)\\
&&\rho_{22}=\alpha\Big(\mu\nu\rho_{11}^{(0)}+\mu^2\rho_{22}^{(0)}+\nu^2\rho_{33}^{(0)}+\mu\nu\rho_{44}^{(0)}\Big)\\
&&\rho_{33}=\alpha\Big(\mu\nu\rho_{11}^{(0)}+\nu^2\rho_{22}^{(0)}+\mu^2\rho_{33}^{(0)}+\mu\nu\rho_{44}^{(0)}\Big)\\
&&\rho_{44}=\alpha\Big(\nu^2\rho_{11}^{(0)}+\mu\nu\rho_{22}^{(0)}+\mu\nu\rho_{33}^{(0)}+\mu^2\rho_{44}^{(0)}\Big)\\
&&\rho_{14}=\alpha\mu^2\rho_{14}^{(0)},\rho_{23}=\alpha\mu^2\rho_{23}^{(0)};~~~\alpha=(\mu+\nu)^{-2}.
     \end{eqnarray}
   The symbols $\mu$ and $\nu$ represent the survival and crosstalk amplitudes respectively. We now move on to show that in spite of decoherence, one may retrieve some error-immune information encoded in the initial state.\\
   \noindent{\textbf{ Retrieval of information:}} The quantity
   \begin{align}
       I = \dfrac{-i\big(\rho_{23}-\rho_{32}\big)}{\rho_{23}+\rho_{32}}=\dfrac{-i\big(\rho_{23}^{(0)}-\rho_{32}^{(0)}\big)}{\rho_{23}^{(0)}+\rho_{32}^{(0)}} \equiv \tan \phi,
   \end{align}
   remains invariant during the noisy evolution.  The invariant $I$ determines the relative phase between the basis states $\ket{\ell,-\ell}$ and $\ket{-\ell,\ell}$ up to a discrete ambiguity. This ambiguity can be removed by determining the signs of real and imaginary part of $\rho_{23}$.

   There are other quantities that remain formally invariant during the noisy evolution,
   \begin{eqnarray}
       &&I_1=\dfrac{\rho_{11}-\rho_{44}}{\rho_{22}-\rho_{33}}=\dfrac{\rho_{11}^{(0)}-\rho_{44}^{(0)}}{\rho_{22}^{(0)}-\rho_{33}^{(0)}},~~~I_2=\dfrac{\rho_{14}+\rho_{41}}{\rho_{14}-\rho_{41}}=\dfrac{\rho_{14}^{(0)}+\rho_{41}^{(0)}}{\rho_{14}^{(0)}-\rho_{41}^{(0)}},\\
       &&I_3=\dfrac{\rho_{14}+\rho_{41}}{\rho_{23}+\rho_{32}}=\dfrac{\rho_{14}^{(0)}+\rho_{41}^{(0)}}{\rho_{23}^{(0)}+\rho_{32}^{(0)}}.
   \end{eqnarray}
 However, since for the state (\ref{Initial_rho0}),   $\rho_{11}^{(0)}=\rho_{44}^{(0)}$ and $\rho_{14}^{(0)} = \rho_{41}^{(0)}=0$, the invariants $I_1, I_2, I_3$ are identically equal to zero.

\subsection{Retrieval of information from a `two-qubit' OAM entangled state through Kolmogorov turbulence}
We next study the example of propagation of an entangled state through atmospheric turbulence.  In \cite{leonhard2015universal}, the effect of atmospheric turbulence, following Kolmogorov model of turbulence, on a two-photon OAM entangled state,
\begin{align}
    \ket{\psi_0} = \dfrac{1}{\sqrt{2}}\Big(\ket{\ell,-\ell}+e^{i\gamma}\ket{-\ell, \ell}\Big),
\end{align}
is simulated for a wide range of $\ell$ ($5-100$), under the assumption that  both the photons undergo the same turbulence effect. The state $\ket{\psi_0}$ has only one free parameter, {\it viz.}, $\gamma$. The effect of turbulence is described by a map whose non-zero elements are determined in \cite{leonhard2015universal} by employing exchange and inversion symmetries.  In \cite{leonhard2015universal}, the final state is determined in the truncated  ordered basis $\{\ket{\ell,\ell},\ket{\ell,-\ell},\ket{-\ell,\ell},\ket{-\ell,-\ell}\}$ to be,
\begin{equation}
      \rho =\frac{1}{2(a+b)^2}\begin{pmatrix}
          2ab&0&0&0\\
          0&a^2+b^2&a^2e^{-i\gamma}&0\\
          0&a^2e^{i\gamma}&a^2+b^2&0\\
          0&0&0&2ab
      \end{pmatrix}
      \label{rho_fin}
\end{equation}
The symbols $a$ and $b$ represent respectively the survival and crosstalk amplitudes. As before, we  show that in spite of atmospheric turbulence, one may identify such quantities that remain invariant.\\
\noindent{\textbf{Retrieval of information:}} 
From equation (\ref{rho_fin}),  we identify the quantity,
\begin{equation}
    I=\frac{-i\big(\rho_{23}-\rho_{32}\big)}{\rho_{23}+\rho_{32}} \equiv \tan \gamma,
\end{equation}
that remains invariant. This quantity determines the relative phase $(\gamma)$ between the states $\ket{\ell,-\ell}$ and $\ket{-\ell,\ell}$ up to a discrete ambiguity.  As in the previous case, one can remove this discrete ambiguity by determining the signs of the real and the imaginary parts of $\rho_{23}$. In this manner, the parameter $\gamma$ can be uniquely determined. Since this is the {\it only unknown parameter} in the initial state, determination of $\gamma$ is equivalent to determining the initial state of the system.
\subsection{Retrieval of information from `three-qubit' OAM entangled state passing through atmospheric turbulence }
We now move  on to an example of propagation of `three-qubit' OAM entangled state through atmospheric turbulence reported in \cite{yan2017effect}. It turns out that in spite of a large number of parameters (see equation (\ref{eq:three-qubit})), it is possible to extract complete information on the state. For this reason, it merits a brief description. 
 
  In \cite{yan2017effect}, the initial state has been assumed to pass through turbulent atmosphere modelled through  Kolmogorov  turbulence.  The authors have assumed a three-photon state initially in the superposition of five modes \cite{yan2017effect},
\begin{align}\label{eq:three-qubit}
    \ket{\psi_0}=\alpha_1\ket{\ell,-\ell,-\ell}+\alpha_2\ket{-\ell, \ell,-\ell}+\alpha_3\ket{-\ell,-\ell,\ell}+\alpha_4\ket{\ell,\ell, \ell}+\alpha_5\ket{-\ell,-\ell,-\ell}.
\end{align}
The information content in the state is carried by bilinears of the form  $\alpha_i\alpha_j^*$ of which eight are independent. Under the  assumption  that all the three photons suffer the same turbulence \cite{yan2017effect}, the authors  arrive at final state which has additional three modes signifying the effect of turbulence.
 The final state in the  truncated ordered basis $\{\ket{\ell,\ell,\ell},\ket{\ell,\ell,-\ell}, \ket{\ell,-\ell,\ell},$ $\ket{\ell,-\ell,-\ell},\ket{-\ell,\ell,\ell},\ket{-\ell,\ell,-\ell},\ket{-\ell,-\ell,\ell},\ket{-\ell,-\ell,-\ell}\}$ is reported to be \cite{yan2017effect},
\begin{align}
    \rho_{\rm out} &= \begin{pmatrix}
    {\bf A} & {\bf B}\\
    {\bf B^{\dagger}} & {\bf C}
    \end{pmatrix},
          \label{Density matrix}
\end{align}
where the block matrices are given by,
\begin{align}
     {\bf A} = \begin{pmatrix}
        G_1 & 0 & 0 & M_{13}\\
         0 & G_2 & 0 & 0\\
         0 & 0 & G_3 & 0\\
         M_3 & 0 & 0 & G_4
         \end{pmatrix},~
         {\bf B} &= \begin{pmatrix}
         0 & M_{14} & M_{15} & M_{16}\\
     0 & 0 & 0 & 0\\
         0 & 0 & 0 & 0\\
         0 & M_1 & M_2 & M_4\end{pmatrix};~
         {\bf C} =\begin{pmatrix} G_5 & 0 & 0 & 0\\
         0 & G_6 & M_6 & M_8\\
         0 & M_{10} & G_7 & M_{12}\\
         0 & M_{18} & M_{19} & G_8\end{pmatrix}.
         \label{Density_matrix_1}
\end{align}
The functional forms of the entries are given explicitly in the appendix (\ref{elements}).
 
\subsubsection{Retrieval of information with invariants}
It follows from equations (\ref{Density matrix}) and (\ref{Density_matrix_1}) that  there are a large number of invariants, totalling to twenty in number (the complete set of the invariants is given in the appendix(\ref{elements})). This far exceeds the number of independent parameters in the initial state  which are only eight  in number. Hence, this provides an excellent example where the modelling employed may be checked self-consistently. We employ the following invariants:
\begin{align}
  &  I_1 =\frac{-i(M_1-M_5)}{M_1+M_5},I_2=\frac{ -i(M_2-M_9)}{M_2+M_9},   I_3=\frac{-i(M_3-M_{13})}{ M_3+M_{13}},I_4=\frac{ -i(M_4-M_{17})}{M_4+M_{17}},\nonumber\\
  &I_5 =\frac{M_1}{M_2},~~~I_6=\frac{M_1}{M_3},~~~I_7=\frac{M_1}{M_4},~~~I_8=\frac{M_2}{M_{10}},
\end{align}
for retrieval of information encoded in the initial state.

\noindent{\textbf{Retrieval of parameters of initial state:}} 
The initial state chosen in (\ref{eq:three-qubit}) has eight independent parameters. The first four invariants, {\it viz.}, $I_1-I_4$ determine the relative phase between the successive basis states respectively, up to discrete ambiguities. These ambiguities, as in the previous cases, can be removed  by employing the signs  of numerators and denominators of these invariants separately. The next four invariants, {\it viz.}, $I_5-I_8$, together with the information of the relative phases between the basis states, provide the magnitude of the co-efficients of the basis states.


    This concludes our discussion of retrieval of information for various simulations studied that are already performed  for propagation of OAM modes in turbulent media. 

 We, now, embark on a study of OAM propagation in weakly tubulence regime. Our purpose is to identify possible universal features. 
 
 \section{Universal features of propagation of OAM modes}
 \label{Weak_turbulence}
\begin{center}
    \begin{figure}[!htb]
\centerline{\includegraphics[scale=0.6]{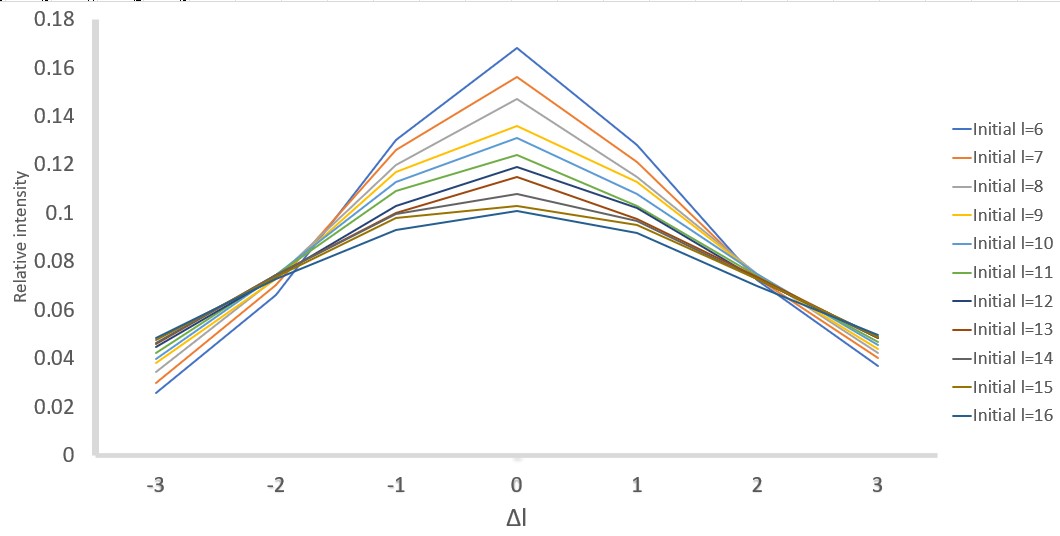}}
\caption{Plot of relative intensities for different values of $\Delta \ell$ for $6 \leq \ell_{\rm in} \leq 16$ for $C_n^2 = 10^{-14} {\rm m}^{-2/3}$. The data has been taken from \cite{anguita2008turbulence}.}
\label{DeltaL}
\end{figure}
\end{center}

There are several studies of OAM propagation in weak atmospheric turbulence regime which are pertinent to our enquiry \cite{anguita2008turbulence, paterson2005atmospheric, tyler2009influence, yang2022influence, zhai2020effects}. In two analytical studies \cite{paterson2005atmospheric, tyler2009influence}, under rather simplified assumptions, it is found that the crosstalk among different modes is independent of the IMI.  In a more extensive numerical simulation \cite{anguita2008turbulence}, it is shown that for turbulence strength of the order of $C_n^2\approx 10^{-15} {\rm m}^{-2/3}$, crosstalk is independent of IMI for small OAM mode indices ($2\leq \ell \leq 7$). We show that the weak dependence on IMI is not restricted to this regime. A careful examination of the data \cite{anguita2008turbulence} reveals that the crosstalk is weakly dependent on IMI for $6 \leq \ell \leq 16$ and $C_n^2=10^{-14}{\rm m}^{-2/3}$.  To demonstrate this, we plot in figure (\ref{DeltaL}) the data given in \cite{anguita2008turbulence} for $C_n^2 = 10^{-14}{\rm m}^{-2/3}$ and  for IMI $6\leq\ell\leq 16$.
  \begin{center}
    \begin{figure}[!htb]
\centerline{\includegraphics[scale=0.5]{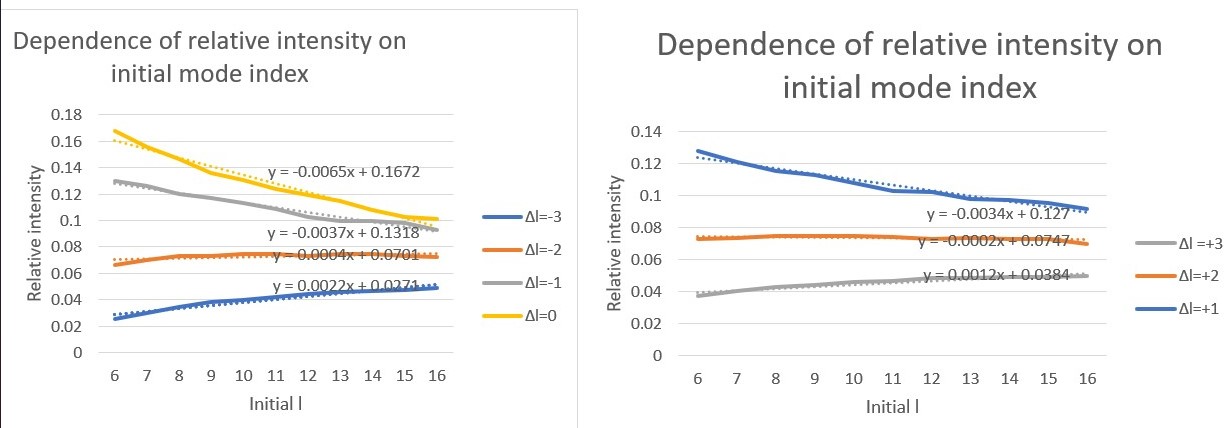}}
\caption{Plot of relative intensities with different initial $\ell$ values for $-3\leq \Delta \ell \leq 3$ for $C_n^2=10^{-14}{\rm m}^{-2/3}$. It becomes clear from this plot that, in this regime, the change in the relative intensities with initial $\ell$ values, characterised by the slope ($\approx 10^{-3}$) is very small for different values of $\Delta \ell$. The data has been taken from \cite{anguita2008turbulence}.}
\label{Dependence}
\end{figure}
\end{center} 
 The feature that concerns us is small change in the relative intensities for a given value of $\Delta \ell$ when IMI changes from $6$ to $16$. For the sake of a better appreciation of this feature,  in figure (\ref{Dependence}), we have plotted  relative intensities with IMI for different values of $\Delta \ell$. The slopes of linear fits for the plots of relative intensities vs. $\Delta \ell$  are of the order of $10^{-3}$, shown in table (\ref{Slope}). The table also contains relative change in the intensities for different values of $\Delta \ell$, which vary from $0.6\%-3\%$, with $\Delta \ell =-3$ being the sole outlier.   The weak dependence on the initial OAM mode  is also argued by universal entanglement decay law \cite{yang2022influence} over a wide range of IMI  ($5\leq \ell \leq 100$).

  Furthermore,  it has been observed in \cite{zhai2020effects} for propagation of `two-qubit' OAM Werner state that coherence survives up to a distance much larger $(50~m)$, in contrast to entanglement which decays much faster $(10~m)$. Similar conclusions have been shown in \cite{wei2017universal} for propagation of `two-qubit' OAM Werner state through atmospheric turbulence.
  \begin{table}[]
      \centering
      \begin{tabular}{||c|c|c|c|c|c|c| c||} 
 \hline
 $\Delta \ell$&$-3$&$-2$&$-1$&$0$&$+1$&$+2$&$+3$\\\hline
  Slope&$0.0022$&$0.0004$&$-0.0037$&$-0.0065$&$-0.0034$&$-0.0002$&$0.0012$\\\hline
 Relative change & & & & & & &\\
 in intensity&$0.085$&$0.006$&$ 0.028$&$0.038$&$0.026$&$ 0.002$&$ 0.037$\\\hline
\end{tabular}
      \caption{Slope of plots of relative intensities  vs. IMI and relative change in intensities for different $\Delta \ell$ values.}
      \label{Slope}
  \end{table}

In summary, the weak dependence shows that idealised crosstalk channel is an excellent approximation to realistic weakly turbulent channels. The  observation that coherence lasts much longer than entanglement provides a  compelling motivation for developing error combating techniques for an IC channel without costly resources such as  multiparty entanglement.

 \subsection{Idealised crosstalk channel}
 \label{crosstalk}
Suppose that the set,
  \begin{align}\label{eq:basis}
     {\cal B} \equiv \{\ket{l_{\rm min}},\cdots ,\ket{l_{\rm max}}\}, 
  \end{align}
   consists of the modes which are employed for information transfer. The effect of an IC channel, given these modes, can be described in terms of the following Kraus operators:
  \begin{equation}\label{eq:Kraus}
      E_{\pm  k}=\sqrt{p_{k}}\sum_{n=l_{\rm min}}^{l_{\rm max}}\ket{n\pm k}\bra{n},
  \end{equation}
   where $p_{k}$ is the probability with which the errors (spillovers) $E_{+k}$ and $E_{-k}$ corrupt a state. In the next section, we identify invariants that allow to transfer error-free information through an IC channel. 
    \section{Identification of invariants for IC channel }
 \label{Framework}
 As is clear from equation (\ref{eq:Kraus}), an IC channel is of the kind in which --the initial and the final states may involve superpositions of $M$ and $N$ modes respectively \cite{paterson2005atmospheric, anguita2008turbulence, tyler2009influence}, i.e., $ {\cal H}^M \xrightarrow[]{{\cal E}_2} {\cal H}^N$. To identify invariants for this channel, we adapt the framework proposed in \cite{Bala22} for noisy evolutions of the kind $ {\cal H}^M \xrightarrow[]{{\cal E}_1} {\cal H}^M$ (${\cal H}^M$ and ${\cal H}^N$ denote $M$ and $N$-- dimensional Hilbert spaces respectively). This adaptation is accomplished by mapping an IC channel to a generalised flip channel acting on a sufficiently higher-dimensional space. 
 In the following, we first briefly recapitulate the invariants for a generalised flip channel.

 \subsection{Invariants for a generalised flip channel}
We briefly recapitulate the method of determination of invariants for a generalised flip channel obtained in \cite{Bala22}. 
A state $\rho$, after passing through a generalised flip channel, changes to a state $\rho'$ as follows:
 \begin{equation}\label{eq:quditflip}
     \rho\rightarrow\rho'= \sum_{r=0}^{N-1} F_r\rho~ F_r^\dagger\equiv \sum_{r=0}^{N-1} p_r X^r\rho~ (X^r)^\dagger,
 \end{equation}
where $p_r$ is the probability with which an error $X^r=\sum_{k=0}^{N-1}\ket{k+r}\bra{k}$ has corrupted a state (here, the summation is modulo $N$). Employing the cyclicity property of trace, we see that the expectation values of operators, 
\begin{equation}
    I_{1}^{(m)}=\langle X^m\rangle,
    \label{Inv1}
\end{equation}
 are invariants and thus the encoded information remain error--free. These invariants belong to {\it the first family}. \\
 Additional invariants are identified by employing the relation $ZX=\omega XZ$, where $Z=\sum_{k=0}^{N-1}\omega^k\ket{k}\bra{k}$, and $\omega$ is the $N^{\rm th}$ root of unity. As shown in \cite{Bala22}, the quantities,
 \begin{equation}
I_{2}^{(m,l)}=\frac{\langle Z^l\rangle}{\langle X^mZ^l\rangle},
\label{Inv2}
 \end{equation}
 also remain invariants.  Hence, information encoded in $I_{2}^{(m, l)}$ also remains immune to error. The invariants $I_2^{(m, l)}$ belong to {\it the second family}.
Equipped with these results, we, first show that the effect of IC channel on a finite-dimensional space can be modelled by a generalised flip channel in a higher-dimensional space and then, move on to identify invariants for an IC channel.

 \subsection{Mapping an IC channel to a generalised flip channel}
 \label{mapping}


To start with, we consider an IC channel  that causes a spillover by one unit. The corresponding error operators are, $E_{\pm 1}\equiv \sqrt{p_{1}}\sum_{k=1}^{M}\ket{k\pm 1}\bra{ k}$.
We, now, consider the state,
\begin{align}\label{eq:relation}
    \ket{\psi} = \sum_{i=1}^{M}\alpha_i\ket{i}\equiv \sum_{i=0}^{M+1}\alpha_i\ket{i};~~~~\alpha_0,\alpha_{M+1}\equiv 0,
\end{align}
that is employed to transfer information.  Then, upto normalisation factors,
\begin{align}\label{eq:crosstalk1}
        {E}_{+1}\ket{\psi} \equiv\sum_{i=1}^{M}\alpha_i\ket{i+1},~~~{ E}_{-1}\ket{\psi} \equiv \sum_{i=1}^{M}\alpha_i\ket{i-1}.
\end{align}

Consider, now, the generalised flip channel acting on ${\cal H}^{M+2}$ via the error operators, $F_{+1}=\sqrt{p_1} X$ and $F_{-1}=F_{+1}^\dagger =\sqrt{p_1}X^\dagger$, where the flip operator is $X=\sum_{k=0}^{M+1}\ket{k+1}\bra{k}$ (with addition modulo $M+2$). Again, upto normalisation factors, the effect of error operators $F_{\pm 1}$ on the state $\ket{\psi}$ is given by, 
\begin{align}\label{eq:flip1}
    F_{+1}\ket{\psi} \equiv \sum_{i=1}^{M}\alpha_i\ket{i+1}, ~~F_{-1}\ket{\psi} \equiv\sum_{i=1}^{M}\alpha_i\ket{i-1},
\end{align}
which is no different from equation (\ref{eq:crosstalk1}), thereby establishing that they have the same effect on the family of states chosen in equation (\ref{eq:relation}).
This argument extends to IC channels that shift a mode by $l$ units provided we employ a generalised flip channel acting on at least $(M+2l)$- dimensional space.

 For an explicit illistration, consider an IC channel that shifts a  mode by $l$ units. The corrrepsonding  error  operators given in equation (\ref{eq:Kraus}) assume the following form,
\begin{align}
    E_{\pm r} = \sqrt{p_r}\sum_{k=l}^{l+M-1}\ket{k\pm  r}\bra{k},~~~~~1\leq r\leq l.
\end{align}
 The initial state would have the form,
 \begin{align}
     \ket{\phi} = \sum_{j=l}^{l+M-1}\beta_j\ket{j}.
 \end{align}
Upto normalisation factors, the effect of these operators on the state $ \ket{\phi}$ is,
\begin{align}\label{eq:crosstalk_l}
     E_{+r}\ket{\phi}\equiv\sum_{j=l}^{l+M-1}\beta_j\ket{j+r},~~~~E_{-r}\ket{\phi}\equiv\sum_{j=l}^{l+M-1}\beta_j\ket{j-r},
\end{align}
 which can be reproduced by the generalised flip channel defined in $(M+2l)$ -- dimensional space with operators,
\begin{align}
F_{+r}=\sqrt{p_r}X^r,~
~~~~~ F_{-r}=F_{+r}^\dagger=\sqrt{p_r}(X^r)^\dagger;~1\leq r\leq l,
\label{GFlip}
\end{align}
where $X^r=\sum_{k=0}^{M+2l-1}\ket{k+r}\bra{k}$ (with addition modulo $M+2l$)  as may be verified easily.
 
 In summary, the effect of a crosstalk error causing a spillover by $l$ units is equal to that of a generalised flip errors in ${\cal H}^{\cal N}$ provided the initial state is chosen appropriately and ${\cal N}\geq M+2l$.

Given the equivalence between the two, the invariants of a generalised flip channel identified  in \cite{Bala22} also serves as invariants for IC channels (given in equations (\ref{Inv1}) and (\ref{Inv2}))  and hence can be used for error--immune information transfer.
In the following section, we describe the scheme that allows for full information retrieval from a state passing through a crosstalk channel that causes a spillover by at most $l$ units.

\section{Complete information retrieval from a state after passing through an IC channel}
\label{information}
In this section, we lay down a scheme that allows for  complete information retrieval from a state by employing the invariants identified in the previous section.  Consider a state involving superposition of $M$ modes, 
\begin{equation}
    \rho=\sum_{i,j=l}^{l+M-1}\rho_{i,j}\ket{i}\bra{j},
\end{equation}
that is employed to transfer information through a crosstalk channel. Suppose that the crosstalk channel is such that it can cause spillover to at most $l$ modes. In that case, the final state, after passing through this channel, takes the form,
 \begin{equation}
    \rho'\equiv \sum_{ k=0}^{l} E_{\pm k}\rho E_{\pm k}^\dagger \equiv \sum_{ k=0}^{l}p_k\rho_{\pm }^{(k)},
    \label{21}
 \end{equation}
where $\rho_{\pm}^{(0)}\equiv\rho$ and $\rho_{\pm }^{(k)}$ are given by,
\begin{align}
   \rho_{\pm }^{(k)}=\sum_{i,j=l}^{l+M-1}\rho_{i,j}\ket{i\pm k}\bra{j\pm k},~~~\forall ~k\in \{1,\cdots, l\}.
    \label{22}
\end{align}
From equations (\ref{21}) and (\ref{22}), it is clear that the final state $\rho'$ has a support in an $(M+2l)$--dimensional Hilbert space.

For the purpose of information retrieval, we first fix the the dimension of the space in which the flip operators are to be defined. This requires us to take into account the following point. The dimesnionality of the space should be  such that the relation,
\begin{align}
\langle X^{M-1}Z^r\rangle_{\rho} =f(\omega,r)\langle X^{M-1}\rangle_{\rho},    
\label{Condition}
\end{align}
holds where $f(\omega, r)$ is a function of only $\omega$ and $r$. 
 Since the initial state $\rho$ involves superposition of $M$ modes, condition (\ref{Condition}) is satisfied iff the operator $X^{M-1}$ is not its own inverse. It holds in  the minimum dimension $2M-1$. This is because in $(2M-1)$--dimensional space, the operator $X$ satisfies the relation, $X^{2M-1}=\mathbb{1}$ leading to $ (X^{M-1})^{-1}=X^M$. On the other hand, since the final state has support in $(M+2l)$ dimensional space, the minimum dimension for information retrieval is given by, 
\begin{equation}\label{eq:dimensionality}
    {\cal N}\equiv{\rm max} [2M-1,M+2l].
\end{equation}

\begin{center}
    \begin{figure}[!htb]
\centerline{\includegraphics[scale=0.35]{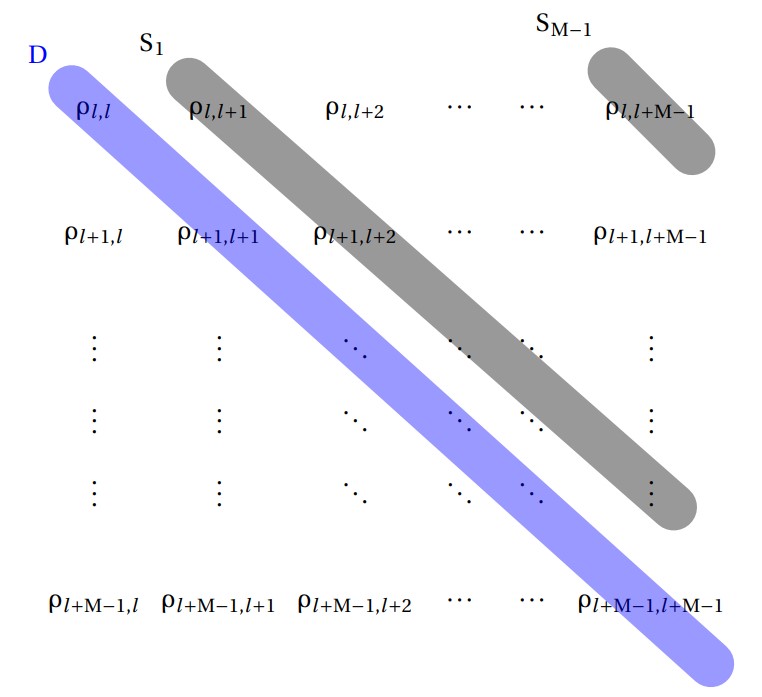}}
\caption{$\rho_{i,j} $ represents the $ij^{\rm th}$ element of the initial density matrix $\rho$, $D$ represents its diagonal, and $S_i$ represents its $i^{\rm th}$ super--diagonal (which we use for convenience).}
\label{state}
\end{figure}
\end{center}

Thus, in the ${\cal N}$--dimensional space, operators $X^r$ and $Z^r$ have the forms:
\begin{align}
    X^r=\sum_{k=0}^{{\cal N}-1}\ket{k+r}\bra{k},~~~~~Z^r=\sum_{k=0}^{{\cal N}-1}\omega^{rk}\ket{k}\bra{k},
\end{align}
where addition is modulo ${\cal N}$ and $\omega$ is ${\cal N}^{\rm th}$ root of identity.
Following the reasoning  in section (\ref{mapping}), the invariants of the first family provide the following information on the initial state $\rho$,
\begin{align}
   I_1^{(r)}= \langle X^r\rangle =\sum_{k=l}^{l+M-1-r}\rho_{k,k+r};~~~~1\leq r\leq M-1.
\end{align}
In a similar manner, the second family of  invariants yields the relation,
\begin{align}
    I_2^{(r,s)}=\frac{\langle Z^s\rangle}{\langle X^rZ^s\rangle}=\frac{\sum_{k=l}^{l+M-1}\omega^{ks}\rho_{k,k}}{\sum_{k=l}^{l+M-1-r}\omega^{ks}\rho_{k,k+r}},~~~~1\leq r,s\leq M-1.
\end{align}
 The steps of the scheme that allows full information retrieval of a state are as follows:
\begin{itemize}
    \item {\bf Step 1:} The first family invariant, $I_1^{(M-1)}$ determines the off--diagonal element $\rho_{l, l+M-1}$.
    \item {\bf Step 2:} The second family of invariants, $$I_2^{(M-1,s)}=\frac{\sum_{k=l}^{l+M-1}\omega^{ks}\rho_{k,k}}{\omega^{ls}\rho_{l,l+M-1}};~~~~1\leq s\leq M-1,$$
\end{itemize}
yield the simultaneous set of $(M-1)$ equations. These equations, together with the trace condition $\sum_{i=l}^{l+M-1}\rho_{i,i}=1$, allow to retrieve each of the diagonal element $\rho_{i,i}.$  \\
It remains to retrieve the other off--diagonal elements for which the following invariants are employed,

 \begin{align}
        I_1^{(p)} =\sum_{k=l}^{l+M-1-p}\rho_{k,k+p}~,~~ I_2^{(p,s)}=\frac{\sum_{k=l}^{l+M-1}\omega^{ks}\rho_{k,k}}{\sum_{k=l}^{l+M-1-p}\omega^{ks}\rho_{k,k+p}},~~~~1\leq s\leq M-1-p,
    \end{align}
    where $p$ can take values from the set $\{1,2,\cdots,M-2\}$. The index $p$ corresponds to the retrieval of off--diagonal elements lying on the line $S_p$ as shown in the figure (\ref{state}). 
    This is explicitly detailed in the table (\ref{Retrieval_1}).

\begin{table}[h!]
\begin{center}
\begin{tabular}{| c| c|c|}
\hline
{\centering\bf Invariants}&{\centering\bf Elements of density matrix}&\\
\hline\hline
 $I_{1}^{(M-1)}$  & $\rho_{i,j};~j-i = M-1$&$S_{M-1}$\\
 \hline
$  I_{2}^{(M-1,s)}, 1\leq s\leq M-1$ & $\rho_{i,i},~ \forall i \in \{l, \cdots, l+M-1\}$&$D $\\
 $I_{1}^{(M-1)},~~\sum_{i=l}^{l+M-1}\rho_{i,i}=1.$ && \\
 \hline

$I_{1}^{(M-2)}, I_{2}^{(M-2, 1)}$ & $\rho_{i, j};~j-i =M-2$&$S_{M-2}$\\
\hline
$I_{1}^{(M-3)},I_{2}^{(M-3,1)}, I_{2}^{(M-3, 2)}$ & $\rho_{i, j},~ j-i =M-3$&$S_{M-3}$\\
\hline
$\vdots$& $\vdots$&$\vdots$ \\
\hline
   $I_{1}^{(1)}, I_{2}^{(1,1)}, I_{2}^{(1,2)}, \cdots, I_{2}^{(1,M-2)}$ & $\rho_{i,j};~j-i =1$&$S_1$\\
    \hline\hline
 \end{tabular}
 \vspace{0.2cm}
\caption{Sets of invariants employed and correspondingly retrieved density matrix elements.}
    \label{Retrieval_1}
    \end{center}
\end{table}
As an illustration,  we have explicitly shown information retrieval through an example of a state initially in superposition of three modes passing through an IC channel causing a spillover by one unit in appendix (\ref{information_qutrit}). 

In the next section, we construct an ancilla-free error-rejecting codes for an IC channel for propagation of OAM modes.
\section{Quantum error rejection code (QERC)}
 \label{rejection}
In this section, we determine the additional constraints to be imposed on the choice of basis modes to construct QERC.  Recall that, a QERC merely detects whether error has corrupted a state or not \cite{wang2004quantum}.  In order to construct a QERC, it is necessary that each error projects the initial state to a space which has no overlap with it. That is,
\begin{align}
    {\cal H}^M \perp {\cal H}^{E},~~~~~{\rm where} ~{\cal H}^{E}\equiv{\cal H}^{E_{+1}} \cup \cdots \cup {\cal H}^{E_{+l}}\cup {\cal H}^{E_{-1}}\cup \cdots \cup {\cal H}^{E_{-l}}.
\end{align}
By virtue of this orthogonality, one may employ an observable,
\begin{equation}
    O=\Pi^M,
\end{equation}
that provides an outcome only if the state is uncorrupted. A more general observable, $O=c_1\Pi^M+c_2\Pi^E,~ c_1 \neq c_2$ can be employed for detection of occurrence of errors. Here, $\Pi^M$ and $\Pi^E$ represent the projections having supports in ${\cal H}^M$ and ${\cal H}^{E}$. If the measurement outcome is $c_1$, then, there is no corruption. On the other hand, the outcome $c_2$ corresponds to the case when the post-measurement state is corrupted and hence can be rejected.

\subsection{QERC for an IC channel that causes a spillover by at most $l$ units}

For an IC channel that shifts a mode by $l$ units, the basis modes should be chosen at a distance,  $\Delta_r=(l+1)$ to preserve the condition of orthogonality of the initial state with all the corrupted states.   Therefore, if an $M$--dimensional state is to be transferred through an IC channel causing a spillover by $l$ units, the following set of basis modes,
\begin{align}
 {\cal B}_r \equiv   \{\ket{l},\ket{l+\Delta_r},\ket{l+2\Delta_r},\cdots,\ket{l+(M-1)\Delta_r}\},
\end{align}
can be employed. Thus, any $M$--dimensional state,
\begin{align}
    \ket{\psi}=\sum_{i=0}^{M-1}\alpha_i\ket{l+i\Delta_r},
\end{align}
can be employed as it remains orthogonal to any of the corrupted states of an IC channel, thus fulfilling the requirement of QERC. For this state, measurement of the observable,
\begin{equation}
    O=\Pi^M=\sum_{j=0}^{M-1}\ket{l+j\Delta_r}\bra{l+j\Delta_r},
\end{equation}
 gives the outcome only when the state is uncorrupted.
 \section{Quantum error correction code (QECC)}
   \label{correction}
   The demand of error correction codes imposes the most stringent condition that the errors projects the initial state onto mutually orthogonal spaces \cite{knill1997theory,suter2016colloquium}. That is, 
   \begin{align}
    {\cal H}^M \perp {\cal H}^{E_{k}},~~{\cal H}^{E_j} \perp {\cal H}^{E_k};~~ j,k\in\{\pm 1,\pm 2,\cdots, \pm l\}.
 \end{align}
 By virtue of this orthogonality,  an observable can be constructed which can distinguish each of these states and hence, each error can be detected. Such an observable is given as,
 \begin{equation}\label{eq:stabiliser_QECC}
     O=c_M\Pi^M+\sum_{k=1}^{l}c_{\pm k}\Pi^{E_{\pm k}},
 \end{equation}
  with distinct eigenvalues $c_M,~c_{\pm k}$ belonging to the respective projections $\Pi^M,~\Pi^{E_{\pm k}}$.

Each outcome of this observable provides information about the subspace to which the post-measurement state belongs. If an error has occurred, its effect can be undone by performing an appropriate transformation on the state.

\begin{center}
    \begin{figure}[!htb]
\centerline{\includegraphics[scale=0.33]{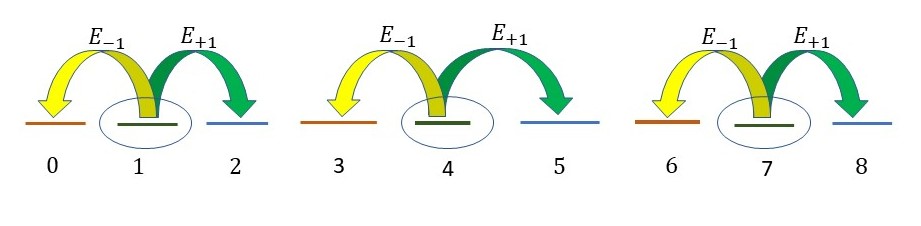}}
\caption{Choice of  modes for a  state passing through a crosstalk channel of one unit. Since the channel causes a crosstalk of one unit, the  modes are to be chosen at an interval of at least three units, so that the error operators, {\it viz.}, ${E}_{\pm 1}$  transform the original state to orthogonal subspaces. }
\label{Choice}
\end{figure}
\end{center}

\noindent{\bf Example: } To start with, we consider an IC channel that causes a spillover by one unit. In this case, the error operators are given as $E_{\pm 1}$. The effect of this crosstalk channel on the basis modes is shown in figure (\ref{Choice}).

For a state involving superposition of three modes, the basis modes  $\{\ket{1},\ket{4},\ket{7}\}$ satisfy the requirement of a QECC. 

Consider a state,
    \begin{align}    \ket{\psi}=\alpha\ket{{1}}+\beta\ket{4}+\gamma\ket{7},
\end{align}
that is employed to transfer information.
 After passing through a crosstalk channel, the final state takes the form,
 \begin{align}     \rho'=p_0\ket{\psi}\bra{\psi}+p_1\big(\ket{\psi_+}\bra{\psi_+}+\ket{\psi_-}\bra{\psi_-}\big).
 \end{align}
 The states $\ket{\psi_\pm} $ have the forms,
\begin{align}
\ket{\psi_+}=\alpha\ket{2}+\beta\ket{5}+\gamma\ket{8},~~
    \ket{\psi_-}=\alpha\ket{0}+\beta\ket{3}+\gamma\ket{6}.
\end{align}
It is to be noted that for this choice of modes, $\langle \psi_{\pm}\ket{\psi} =0$ and $\bra{\psi_+}\psi_-\rangle=0$. Detection of errors can be performed by measuring the stabiliser given in equation (\ref{eq:stabiliser_QECC}) for which the projections are:
\begin{align}
&\Pi^M = \ket{{1}}\bra{{1}}+\ket{{4}}\bra{{4}}+\ket{{7}}\bra{{7}},~\Pi^{E_{+ 1}}=\ket{2}\bra{2}+\ket{5}\bra{5}+\ket{8}\bra{8},\\
   & \Pi^{E_{- 1}}=\ket{0}\bra{0}+\ket{3}\bra{3}+\ket{6}\bra{6}.
\end{align}
Since each outcome of the stabiliser provides information about occurrence of a particular error, effect of which can be undone by performing an appropriate transformation. The measurement outcomes, corresponding errors and required transformations for error correction have  been given in table (\ref{tab:qubit1}).
\begin{table}[h!]
\begin{center}
\begin{tabular}{| c| c |c|}
\hline
\multirow{3}{2.5cm}{{\centering\bf Measurement outcomes}}&\multirow{3}{2cm}{\centering\bf Error operator} &\multirow{3}{2.5cm}{{\bf Transformation}}\\
 & & \\
 &  &\\
\hline\hline
 $c_M$ 
&$E_0$& No error \\\hline
$c_{-1}$&${E}_{-1}$& $\ket{{1}}\bra{ {0}}+\ket{{4}}\bra{ {3}}+\ket{{7}}\bra{ {6}}$\\\hline
$c_{+1}$& ${E}_{+1}$&  $\ket{{1}}\bra{ {2}}+\ket{4}\bra{ {5}}+\ket{{7}}\bra{{8}}$ \\ 
\hline
 \end{tabular}
 \vspace{0.2cm}
\caption{ Outcomes of stabiliser measurement, error detection and corresponding transformations for its correction.}
    \label{tab:qubit1}
    \end{center}
\end{table}

For the sake of generalisation, in appendix (\ref{Identification}), we have constructed a QECC that allows to retrieve state after passing through an IC channel that causes a spillover by $l$ units. 

 \section{Conclusion}
 \label{conclusion}
 In summary, we have accomplished two tasks in this paper: (i) information retrieval when entangled biphoton and three-photon OAM states propagate through atmospheric turbulence and oceanic turbulence from  analytical and simulation data.  (ii) We exploit the weak dependence of crossover probability on initial OAM mode index in weak atmospheric turbulence to demonstrate  retrieval of information from a state passing through an IC channel if the receiver has a capability of performing measurements in an arbitrarily large subspace of OAM modes.   We have also constructed quantum error-correcting and error-rejecting codes for an IC channel, which could be potentially useful.

\begin{appendices}
\section{Complete list of invariants and the elements of final density matrix for a `three-qubit' entangled state}\label{elements}
{\it Elements of final density matrix for a `three-qubit' entangled state:} The elements of the final state are functions of initial state parameters and survival $(a)$ and crosstalk amplitudes $(b)$ which are reported to be \cite{yan2017effect}:
\begin{align}
    G_1 &=\vert\alpha_4\vert^2a^3+\Big(\vert\alpha_1\vert^2+\vert\alpha_2\vert^2+\vert\alpha_3\vert^2\Big)ab^2+\vert\alpha_5\vert^2b^3,\nonumber\\
    G_2 &=\Big(\vert\alpha_1\vert^2+\vert\alpha_4\vert^2+\vert\alpha_2\vert^2\Big)a^2b+\vert\alpha_5\vert^2ab^2+\vert\alpha_3\vert^2b^3,\nonumber\\
    G_3 &=\Big(\vert\alpha_1\vert^2+\vert\alpha_3\vert^2+\vert\alpha_4\vert^2\Big)a^2b+\vert\alpha_2\vert^2b^3+\vert\alpha_5\vert^2ab^2,\nonumber\\
    G_4 &= \vert\alpha_1\vert^2a^3+\Big(\vert\alpha_2\vert^2+\vert\alpha_3\vert^2+\vert\alpha_4\vert^2\Big)ab^2+\vert\alpha_5\vert^2a^2b,\nonumber\\
    G_5 &=\vert\alpha_1\vert^2b^3+\Big(\vert\alpha_2\vert^2+\vert\alpha_3\vert^2+\vert\alpha_4\vert^2\Big)a^2b+\vert\alpha_5\vert^2ab^2,\\
    G_6 &= \vert \alpha_2\vert^2a^3+\Big(\vert \alpha_1\vert^2+\vert \alpha_3\vert^2+\vert \alpha_4\vert^2\Big)ab^2+\vert\alpha_5\vert^2 a^2b,\nonumber\\
    G_7 &= \vert \alpha_3\vert^2 a^3 +\Big(\vert \alpha_1\vert^2 +\vert\alpha_2\vert^2 +\vert\alpha_4\vert^2\Big)ab^2+\vert\alpha_5\vert^2a^2b,\nonumber\\
    G_8 &= \vert\alpha_5\vert^2a^3+\Big(\vert\alpha_1\vert^2+\vert\alpha_2\vert^2+\vert\alpha_3\vert^2\Big)a^2b+\vert\alpha_4\vert^2 b^3,\nonumber\\
    M_1 &=M_5^{*}=\alpha_1\alpha_2^*a^3, M_2 =M_9^{*}=\alpha_1\alpha_3^*a^3,~~M_3 =M_{13}^{*}=\alpha_1\alpha_4^*a^3, M_4 =M_{17}^*= \alpha_1\alpha_5^*a^3\nonumber\\
  M_6 &=M_{10}^{*}=\alpha_2\alpha_3^*a^3,~~~
M_7 =M_{14}^{*}=\alpha_2\alpha_4^*a^3 , M_8 =M_{18}^{*}=\alpha_2\alpha_5^*a^3\nonumber\\
 M_{11} &=M_{15}^{*}=\alpha_3\alpha_4^*a^3, M_{12} =M_{19}^{*}=\alpha_3\alpha_5^*a^3\nonumber\\
 M_{16} &=M_{20}^{*}=\alpha_4\alpha_5^*a^3
\end{align}
\noindent{\it Complete list of invariants:} For the sake of completeness, the invariants are given as follows:
\begin{align}
    I_1 &=\frac{i(M_1-M_5)}{M_1+M_5}; I_2=\frac{M_2+M_9}{M_1+M_5};I_3=\frac{i(M_2-M_9)}{M_2+M_9};I_4=\frac{i(M_3-M_{13})}{M_3+M_{13}};\nonumber\\
     I_5 &=\frac{M_3+M_{13}}{M_1+M_5}; I_6=\frac{i(M_4-M_{17})}{M_4+M_{17}};I_7=\frac{M_4+M_{17}}{M_1+M_5};I_8=\frac{i(M_6-M_{10})}{M_{6}+M_{10}};\nonumber\\
     I_9 &=\frac{M_6+M_{10}}{M_{1}+M_5};I_{10}=\frac{i(M_7-M_{14})}{M_7+M_{14}}; I_{11}=\frac{M_7+M_{14}}{M_{1}+M_5}; I_{12}=\frac{i(M_8-M_{18})}{M_8+M_{18}};\nonumber\\
        I_{13} &=\frac{M_8+M_{18}}{M_{1}+M_5};I_{14}=\frac{i(M_{11}-M_{15})}{M_{11}+M_{15}}; I_{15}=\frac{M_{11}+M_{15}}{M_{1}+M_5}; I_{16}=\frac{i(M_{12}-M_{19})}{M_{12}+M_{19}};\nonumber\\
        I_{17} &=\frac{M_{12}+M_{19}}{M_{1}+M_{5}}; I_{18}=\frac{i(M_{16}-M_{20})}{M_{16}+M_{20}}; I_{19}=\frac{M_{16}+M_{20}}{M_1+M_5}; I_{20}=\frac{G_4-G_6}{G_6-G_7}.
\end{align}
%
\section{Example: Information retrieval from a state initially in superposition of three modes passing through an IC channel causing a spillover by one unit}
\label{information_qutrit}
This appendix acts as a pedagogic illustration. In this, we show information retrieval from a state initially in superposition of three modes passing through an IC channel causing crosstalk by one unit. 
 The error operators of a crosstalk channel causing a spillover by one unit are given by,
\begin{align}
E_0= \sqrt{p_0}\sum_{k=1}^3\ket{k}\bra{k},~~    E_{\pm 1} \equiv\sqrt{p_1} \sum_{k=1}^3\ket{k\pm 1}\ket{k}.
\end{align} 
Suppose that a state,
\begin{align}
     \rho 
     &=\sum_{i, j=1}^3\rho_{i,j}\ket{i}\bra{j},
 \end{align}
is employed to transfer information. The resulting state, $\rho'$, after passing through this channel is given by,
 \begin{align}
     \rho'&\equiv E_{0}\rho E_{0}^\dagger+E_{+1}\rho E_{+1}^\dagger+E_{-1}\rho E_{-1}^\dagger\equiv p_0\rho +p_1\rho_{+}^{(1)}+p_1\rho_{-}^{(1)}.
 \end{align}
 The states $\rho_{\pm}^{(1)}$ would have the forms,
 \begin{align}
     \rho_{\pm}^{(1)} = \sum_{i, j=1}^3\rho_{i, j}\ket{i\pm 1}\bra{j \pm 1}.
 \end{align}
Since the corresponding final state $\rho'$ involves superpositions of five modes, we define the corresponding flip and phase operators as $X  = \sum_{j=0}^4\ket{j+1}\bra{j}$ and $Z=\sum_{j=0}^4\omega^j\ket{j}\bra{j}$ (with addition modulo 5 and $\omega$ is fifth root of unity).

The  invariants of first family are found to be,
\begin{equation}
    I_1^{(1)}=\langle X\rangle =\rho_{1,2}+\rho_{2,3};~~~~I_1^{(2)}=\langle X^2\rangle =\rho_{1,3}.
\end{equation}
Similarly, the  invariants of second family are found to be,
\begin{eqnarray}
\label{eq:inv1}
    &I_2^{(1,1)}=\frac{\langle Z\rangle}{\langle XZ\rangle}=\frac{\omega\rho_{1,1}+\omega^2\rho_{2,2}+\omega^3\rho_{3,3}}{\omega\rho_{1,2}+\omega^2\rho_{2,3}},\nonumber\\
    & I_2^{(1,2)}=\frac{\langle Z^2\rangle}{\langle XZ^2\rangle}=\frac{\omega^2\rho_{1,1}+\omega^4\rho_{2,2}+\omega^6\rho_{3,3}}{\omega^2\rho_{1,2}+\omega^4\rho_{2,3}}\\
    \label{eq:inv2}
     &I_2^{(2,1)}=\frac{\langle Z\rangle}{\langle X^2Z\rangle}=\frac{\omega\rho_{1,1}+\omega^2\rho_{2,2}+\omega^3\rho_{3,3}}{\omega\rho_{1,3}},\nonumber\\ &I_2^{(2,2)}=\frac{\langle Z^2\rangle}{\langle X^2Z^2\rangle}=\frac{\omega^2\rho_{1,1}+\omega^4\rho_{2,2}+\omega^6\rho_{3,3}}{\omega^2\rho_{1,3}}.
\end{eqnarray}
It is important to note that the numerators and denominators in equations (\ref{eq:inv1}) and (\ref{eq:inv2}) have same dependence on the channel parameters $p_0$ and $p_1$, that is why their ratios, i.e., $I_2^{(1, 1)}, I_2^{(1, 2)}, I_2^{(2, 1)}$ and $I_2^{(2, 2)}$ are independent of them. Given these invariants, we next, enumerate the steps to retrieve full information in the state $\rho$ as follows:
\begin{itemize}
    \item {\bf Step 1:} The invariant $I_1^{(2)}$  determines the off--diagonal density matrix element $\rho_{1,3}.$
    \item {\bf Step 2:} Having knowledge of $\rho_{1,3}$, we next employ the invariants $I_2^{(2,1)}$ and $I_2^{(2,2)}$, together with the trace condition $\rho_{1,1}+\rho_{2,2}+\rho_{3,3}=1$, to find the values of  the diagonal elements of the density matrix $\rho$, i.e., $\rho_{1,1},\rho_{2,2}$, and $\rho_{3,3}$.
    \item {\bf Step 3:} Equipped with the diagonal elements of the density matrix $\rho$, we employ the invariants $I_2^{(1,1)}$ and $I_2^{(1,2)}$ to retrieve the off--diagonal density matrix elements $\rho_{1,2}$ and $\rho_{2,3}$.
\end{itemize}
In this manner, we have determined diagonal as well as off--diagonal elements of the density matrix $\rho$. Since the equations get undetermined if $\rho_{1, 3} =0$, it is necessary that the initial state should be prepared accordingly.
\section{QECC for IC channel causing spillover by $l$ units}
\label{Identification}
In section (\ref{correction}), we have constructed a QECC for a state involving superposition of three modes passing through an IC channel causing a spillover by one unit. In this appendix, we generalise it by constructing a QECC for an IC channel causing spillover by $l$ units. For an IC channel causing a spillover by $l$ units, the consecutive basis modes should be chosen at a distance of $\Delta=(2l+1)$. 
So, for an $M$--dimensional system, the following set of basis modes can be employed,
\begin{align}
{\cal B}_c \equiv \{\ket{l},\ket{l+\Delta},\cdots,
\ket{l+(M-1)\Delta}\}.  
\end{align}
The most general $M$--dimensional pure state employing these basis modes is given as,
\begin{align}
    \ket{\psi}\equiv&\sum_{j=0}^{M-1}\alpha_j\ket{l+j\Delta}.
\end{align}
The projections, for basis modes ${\cal B}_c$, that determine the stabiliser given in equation (\ref{eq:stabiliser_QECC})  are,
\begin{align}
   & \Pi^M \equiv \sum_{j=0}^{M-1}\ket{l+j\Delta}\bra{l+j\Delta};~\Delta=2l+1,\\
   &\Pi^{E_{\pm k}} \equiv \sum_{j=0}^{M-1}\ket{l\pm k+j\Delta}\bra{l\pm k+j\Delta},~~~~1\leq k\leq l,
\end{align}
Depending on the outcome of stabiliser, one may apply an appropriate transformation on the state as given in the table (\ref{tab:qecc}).

\begin{table}[h!]
\begin{center}
\begin{tabular}{| c| c |c|}
\hline
\multirow{3}{2.5cm}{{\centering\bf Measurement outcomes}}&\multirow{3}{2cm}{\centering\bf Error} &\multirow{3}{2.5cm}{{\bf Transformation}}\\
 & & \\
 &  &\\
\hline\hline
 $c_0$ 
&$E_0$& No error \\
\hline
$c_{\pm 1}$&${E}_{\pm 1}$& $\sum_{j=0}^{M-1}\ket{l+j\Delta}\bra{l\pm 1+j\Delta}$\\
\hline
$c_{\pm 2}$&${E}_{\pm 2}$& $\sum_{j=0}^{M-1}\ket{l+j\Delta}\bra{l\pm 2+j\Delta}$\\
\hline
$\cdot$&$\cdot$&$\cdot$\\
$\cdot$&$\cdot$&$\cdot$\\
$\cdot$&$\cdot$&$\cdot$\\
\hline
$c_{\pm l}$&${E}_{\pm l}$& $\sum_{j=0}^{M-1}\ket{l+j\Delta}\bra{l\pm l+j\Delta}$\\
\hline

 \end{tabular}
 \vspace{0.2cm}
\caption{Measurement outcomes of the stabiliser, detected error and the corresponding transformation.}
    \label{tab:qecc}
    \end{center}
\end{table}

 \end{appendices}

 




\begin{thebibliography}{30}
\ifx \bisbn   \undefined \def \bisbn  #1{ISBN #1}\fi
\ifx \binits  \undefined \def \binits#1{#1}\fi
\ifx \bauthor  \undefined \def \bauthor#1{#1}\fi
\ifx \batitle  \undefined \def \batitle#1{#1}\fi
\ifx \bjtitle  \undefined \def \bjtitle#1{#1}\fi
\ifx \bvolume  \undefined \def \bvolume#1{\textbf{#1}}\fi
\ifx \byear  \undefined \def \byear#1{#1}\fi
\ifx \bissue  \undefined \def \bissue#1{#1}\fi
\ifx \bfpage  \undefined \def \bfpage#1{#1}\fi
\ifx \blpage  \undefined \def \blpage #1{#1}\fi
\ifx \burl  \undefined \def \burl#1{\textsf{#1}}\fi
\ifx \doiurl  \undefined \def \doiurl#1{\url{https://doi.org/#1}}\fi
\ifx \betal  \undefined \def \betal{\textit{et al.}}\fi
\ifx \binstitute  \undefined \def \binstitute#1{#1}\fi
\ifx \binstitutionaled  \undefined \def \binstitutionaled#1{#1}\fi
\ifx \bctitle  \undefined \def \bctitle#1{#1}\fi
\ifx \beditor  \undefined \def \beditor#1{#1}\fi
\ifx \bpublisher  \undefined \def \bpublisher#1{#1}\fi
\ifx \bbtitle  \undefined \def \bbtitle#1{#1}\fi
\ifx \bedition  \undefined \def \bedition#1{#1}\fi
\ifx \bseriesno  \undefined \def \bseriesno#1{#1}\fi
\ifx \blocation  \undefined \def \blocation#1{#1}\fi
\ifx \bsertitle  \undefined \def \bsertitle#1{#1}\fi
\ifx \bsnm \undefined \def \bsnm#1{#1}\fi
\ifx \bsuffix \undefined \def \bsuffix#1{#1}\fi
\ifx \bparticle \undefined \def \bparticle#1{#1}\fi
\ifx \barticle \undefined \def \barticle#1{#1}\fi
\bibcommenthead
\ifx \bconfdate \undefined \def \bconfdate #1{#1}\fi
\ifx \botherref \undefined \def \botherref #1{#1}\fi
\ifx \url \undefined \def \url#1{\textsf{#1}}\fi
\ifx \bchapter \undefined \def \bchapter#1{#1}\fi
\ifx \bbook \undefined \def \bbook#1{#1}\fi
\ifx \bcomment \undefined \def \bcomment#1{#1}\fi
\ifx \oauthor \undefined \def \oauthor#1{#1}\fi
\ifx \citeauthoryear \undefined \def \citeauthoryear#1{#1}\fi
\ifx \endbibitem  \undefined \def \endbibitem {}\fi
\ifx \bconflocation  \undefined \def \bconflocation#1{#1}\fi
\ifx \arxivurl  \undefined \def \arxivurl#1{\textsf{#1}}\fi
\csname PreBibitemsHook\endcsname

\bibitem{Bennett84}
\begin{botherref}
\oauthor{\bsnm{Bennett}, \binits{C.}},
\oauthor{\bsnm{Brassard}, \binits{G.}}:
Quantum cryptography: public key distribution and coin tossing.
Proc. IEEE Int. Conf. on Comp. Sys. Signal Process (ICCSSP),
175
(1984)
\end{botherref}
\endbibitem

\bibitem{Ekert91}
\begin{barticle}
\bauthor{\bsnm{Ekert}, \binits{A.K.}}:
\batitle{Quantum cryptography based on bell's theorem}.
\bjtitle{Phys. Rev. Lett.}
\bvolume{67},
\bfpage{661}--\blpage{663}
(\byear{1991}).
\doiurl{10.1103/PhysRevLett.67.661}
\end{barticle}
\endbibitem

\bibitem{Bennett93}
\begin{barticle}
\bauthor{\bsnm{Bennett}, \binits{C.H.}},
\bauthor{\bsnm{Brassard}, \binits{G.}},
\bauthor{\bsnm{Cr{\'e}peau}, \binits{C.}},
\bauthor{\bsnm{Jozsa}, \binits{R.}},
\bauthor{\bsnm{Peres}, \binits{A.}},
\bauthor{\bsnm{Wootters}, \binits{W.K.}}:
\batitle{Teleporting an unknown quantum state via dual classical and
  einstein-podolsky-rosen channels}.
\bjtitle{Physical review letters}
\bvolume{70}(\bissue{13}),
\bfpage{1895}
(\byear{1993})
\end{barticle}
\endbibitem

\bibitem{Bennett2001remote}
\begin{barticle}
\bauthor{\bsnm{Bennett}, \binits{C.H.}},
\bauthor{\bsnm{DiVincenzo}, \binits{D.P.}},
\bauthor{\bsnm{Shor}, \binits{P.W.}},
\bauthor{\bsnm{Smolin}, \binits{J.A.}},
\bauthor{\bsnm{Terhal}, \binits{B.M.}},
\bauthor{\bsnm{Wootters}, \binits{W.K.}}:
\batitle{Remote state preparation}.
\bjtitle{Physical Review Letters}
\bvolume{87}(\bissue{7}),
\bfpage{077902}
(\byear{2001})
\end{barticle}
\endbibitem

\bibitem{steane1996simple}
\begin{barticle}
\bauthor{\bsnm{Steane}, \binits{A.M.}}:
\batitle{Simple quantum error-correcting codes}.
\bjtitle{Physical Review A}
\bvolume{54}(\bissue{6}),
\bfpage{4741}
(\byear{1996})
\end{barticle}
\endbibitem

\bibitem{schumacher2002approximate}
\begin{barticle}
\bauthor{\bsnm{Schumacher}, \binits{B.}},
\bauthor{\bsnm{Westmoreland}, \binits{M.D.}}:
\batitle{Approximate quantum error correction}.
\bjtitle{Quantum Information Processing}
\bvolume{1}(\bissue{1}),
\bfpage{5}--\blpage{12}
(\byear{2002})
\end{barticle}
\endbibitem

\bibitem{zanardi1997error}
\begin{barticle}
\bauthor{\bsnm{Zanardi}, \binits{P.}},
\bauthor{\bsnm{Rasetti}, \binits{M.}}:
\batitle{Error avoiding quantum codes}.
\bjtitle{Modern Physics Letters B}
\bvolume{11}(\bissue{25}),
\bfpage{1085}--\blpage{1093}
(\byear{1997})
\end{barticle}
\endbibitem

\bibitem{layden2019ancilla}
\begin{barticle}
\bauthor{\bsnm{Layden}, \binits{D.}},
\bauthor{\bsnm{Zhou}, \binits{S.}},
\bauthor{\bsnm{Cappellaro}, \binits{P.}},
\bauthor{\bsnm{Jiang}, \binits{L.}}:
\batitle{Ancilla-free quantum error correction codes for quantum metrology}.
\bjtitle{Physical review letters}
\bvolume{122}(\bissue{4}),
\bfpage{040502}
(\byear{2019})
\end{barticle}
\endbibitem

\bibitem{lidar1998decoherence}
\begin{barticle}
\bauthor{\bsnm{Lidar}, \binits{D.A.}},
\bauthor{\bsnm{Chuang}, \binits{I.L.}},
\bauthor{\bsnm{Whaley}, \binits{K.B.}}:
\batitle{Decoherence-free subspaces for quantum computation}.
\bjtitle{Physical Review Letters}
\bvolume{81}(\bissue{12}),
\bfpage{2594}
(\byear{1998})
\end{barticle}
\endbibitem

\bibitem{viola1999dynamical}
\begin{barticle}
\bauthor{\bsnm{Viola}, \binits{L.}},
\bauthor{\bsnm{Knill}, \binits{E.}},
\bauthor{\bsnm{Lloyd}, \binits{S.}}:
\batitle{Dynamical decoupling of open quantum systems}.
\bjtitle{Physical Review Letters}
\bvolume{82}(\bissue{12}),
\bfpage{2417}
(\byear{1999})
\end{barticle}
\endbibitem

\bibitem{byrd2004overview}
\begin{barticle}
\bauthor{\bsnm{Byrd}, \binits{M.S.}},
\bauthor{\bsnm{Wu}, \binits{L.-A.}},
\bauthor{\bsnm{Lidar}, \binits{D.A.}}:
\batitle{Overview of quantum error prevention and leakage elimination}.
\bjtitle{journal of modern optics}
\bvolume{51}(\bissue{16-18}),
\bfpage{2449}--\blpage{2460}
(\byear{2004})
\end{barticle}
\endbibitem

\bibitem{proietti2021experimental}
\begin{barticle}
\bauthor{\bsnm{Proietti}, \binits{M.}},
\bauthor{\bsnm{Ho}, \binits{J.}},
\bauthor{\bsnm{Grasselli}, \binits{F.}},
\bauthor{\bsnm{Barrow}, \binits{P.}},
\bauthor{\bsnm{Malik}, \binits{M.}},
\bauthor{\bsnm{Fedrizzi}, \binits{A.}}:
\batitle{Experimental quantum conference key agreement}.
\bjtitle{Science Advances}
\bvolume{7}(\bissue{23}),
\bfpage{0395}
(\byear{2021})
\end{barticle}
\endbibitem

\bibitem{Bala22}
\begin{botherref}
\oauthor{\bsnm{Bala}, \binits{R.}},
\oauthor{\bsnm{Asthana}, \binits{S.}},
\oauthor{\bsnm{Ravishankar}, \binits{V.}}:
Combating quantum errors: an integrated approach.
Submitted to Scientific Reports
(2022)
{\href{https://arxiv.org/abs/2208.04555}{{arXiv:2208.04555}}}
{[quant-ph]}
\end{botherref}
\endbibitem

\bibitem{mirhosseini2015high}
\begin{barticle}
\bauthor{\bsnm{Mirhosseini}, \binits{M.}},
\bauthor{\bsnm{Maga{\~n}a-Loaiza}, \binits{O.S.}},
\bauthor{\bsnm{O’Sullivan}, \binits{M.N.}},
\bauthor{\bsnm{Rodenburg}, \binits{B.}},
\bauthor{\bsnm{Malik}, \binits{M.}},
\bauthor{\bsnm{Lavery}, \binits{M.P.}},
\bauthor{\bsnm{Padgett}, \binits{M.J.}},
\bauthor{\bsnm{Gauthier}, \binits{D.J.}},
\bauthor{\bsnm{Boyd}, \binits{R.W.}}:
\batitle{High-dimensional quantum cryptography with twisted light}.
\bjtitle{New Journal of Physics}
\bvolume{17}(\bissue{3}),
\bfpage{033033}
(\byear{2015})
\end{barticle}
\endbibitem

\bibitem{wang2020satellite}
\begin{barticle}
\bauthor{\bsnm{Wang}, \binits{Z.}},
\bauthor{\bsnm{Malaney}, \binits{R.}},
\bauthor{\bsnm{Burnett}, \binits{B.}}:
\batitle{Satellite-to-earth quantum key distribution via orbital angular
  momentum}.
\bjtitle{Physical Review Applied}
\bvolume{14}(\bissue{6}),
\bfpage{064031}
(\byear{2020})
\end{barticle}
\endbibitem

\bibitem{li2020deterministic}
\begin{barticle}
\bauthor{\bsnm{Li}, \binits{S.}},
\bauthor{\bsnm{Pan}, \binits{X.}},
\bauthor{\bsnm{Ren}, \binits{Y.}},
\bauthor{\bsnm{Liu}, \binits{H.}},
\bauthor{\bsnm{Yu}, \binits{S.}},
\bauthor{\bsnm{Jing}, \binits{J.}}:
\batitle{Deterministic generation of orbital-angular-momentum multiplexed
  tripartite entanglement}.
\bjtitle{Physical Review Letters}
\bvolume{124}(\bissue{8}),
\bfpage{083605}
(\byear{2020})
\end{barticle}
\endbibitem

\bibitem{erhard2020advances}
\begin{barticle}
\bauthor{\bsnm{Erhard}, \binits{M.}},
\bauthor{\bsnm{Krenn}, \binits{M.}},
\bauthor{\bsnm{Zeilinger}, \binits{A.}}:
\batitle{Advances in high-dimensional quantum entanglement}.
\bjtitle{Nature Reviews Physics}
\bvolume{2}(\bissue{7}),
\bfpage{365}--\blpage{381}
(\byear{2020})
\end{barticle}
\endbibitem
\bibitem{bachmann2019universal}
\begin{barticle}
\bauthor{\bsnm{Bachmann}, \binits{D.}},
\bauthor{\bsnm{Shatokhin}, \binits{V.N.}},
\bauthor{\bsnm{Buchleitner}, \binits{A.}}:
\batitle{Universal entanglement decay of photonic orbital angular momentum
  qubit states in atmospheric turbulence: an analytical treatment}.
\bjtitle{Journal of Physics A: Mathematical and Theoretical}
\bvolume{52}(\bissue{40}),
\bfpage{405303}
(\byear{2019})
\end{barticle}
\endbibitem

\bibitem{zhai2020effects}
\begin{barticle}
\bauthor{\bsnm{Zhai}, \binits{S.}},
\bauthor{\bsnm{Zhu}, \binits{Y.}},
\bauthor{\bsnm{Zhang}, \binits{Y.}},
\bauthor{\bsnm{Hu}, \binits{Z.}}:
\batitle{Effects of oceanic turbulence on orbital angular momenta of optical
  communications}.
\bjtitle{Journal of Marine Science and Engineering}
\bvolume{8}(\bissue{11}),
\bfpage{869}
(\byear{2020})
\end{barticle}
\endbibitem
\bibitem{anguita2008turbulence}
\begin{barticle}
\bauthor{\bsnm{Anguita}, \binits{J.A.}},
\bauthor{\bsnm{Neifeld}, \binits{M.A.}},
\bauthor{\bsnm{Vasic}, \binits{B.V.}}:
\batitle{Turbulence-induced channel crosstalk in an orbital angular
  momentum-multiplexed free-space optical link}.
\bjtitle{Applied optics}
\bvolume{47}(\bissue{13}),
\bfpage{2414}--\blpage{2429}
(\byear{2008})
\end{barticle}
\endbibitem
\bibitem{Bachmann_2019}
\begin{barticle}
\bauthor{\bsnm{Bachmann}, \binits{D.}},
\bauthor{\bsnm{Shatokhin}, \binits{V.N.}},
\bauthor{\bsnm{Buchleitner}, \binits{A.}}:
\batitle{Universal entanglement decay of photonic orbital angular momentum
  qubit states in atmospheric turbulence: an analytical treatment}.
\bjtitle{Journal of Physics A: Mathematical and Theoretical}
\bvolume{52}(\bissue{40}),
\bfpage{405303}
(\byear{2019}).
\doiurl{10.1088/1751-8121/ab3f3c}
\end{barticle}
\endbibitem

\bibitem{leonhard2015universal}
\begin{barticle}
\bauthor{\bsnm{Leonhard}, \binits{N.D.}},
\bauthor{\bsnm{Shatokhin}, \binits{V.N.}},
\bauthor{\bsnm{Buchleitner}, \binits{A.}}:
\batitle{Universal entanglement decay of photonic-orbital-angular-momentum
  qubit states in atmospheric turbulence}.
\bjtitle{Physical Review A}
\bvolume{91}(\bissue{1}),
\bfpage{012345}
(\byear{2015})
\end{barticle}
\endbibitem

\bibitem{yan2017effect}
\begin{barticle}
\bauthor{\bsnm{Yan}, \binits{X.}},
\bauthor{\bsnm{Zhang}, \binits{P.-F.}},
\bauthor{\bsnm{Zhang}, \binits{J.-H.}},
\bauthor{\bsnm{Feng}, \binits{X.-X.}},
\bauthor{\bsnm{Qiao}, \binits{C.-H.}},
\bauthor{\bsnm{Fan}, \binits{C.-Y.}}:
\batitle{Effect of atmospheric turbulence on entangled orbital angular momentum
  three-qubit state}.
\bjtitle{Chinese Physics B}
\bvolume{26}(\bissue{6}),
\bfpage{064202}
(\byear{2017})
\end{barticle}
\endbibitem
\bibitem{paterson2005atmospheric}
\begin{barticle}
\bauthor{\bsnm{Paterson}, \binits{C.}}:
\batitle{Atmospheric turbulence and orbital angular momentum of single photons
  for optical communication}.
\bjtitle{Physical review letters}
\bvolume{94}(\bissue{15}),
\bfpage{153901}
(\byear{2005})
\end{barticle}
\endbibitem

\bibitem{tyler2009influence}
\begin{barticle}
\bauthor{\bsnm{Tyler}, \binits{G.A.}},
\bauthor{\bsnm{Boyd}, \binits{R.W.}}:
\batitle{Influence of atmospheric turbulence on the propagation of quantum
  states of light carrying orbital angular momentum}.
\bjtitle{Optics letters}
\bvolume{34}(\bissue{2}),
\bfpage{142}--\blpage{144}
(\byear{2009})
\end{barticle}
\endbibitem
\bibitem{wei2017universal}
\begin{barticle}
\bauthor{\bsnm{Wei}, \binits{M.-S.}},
\bauthor{\bsnm{Wang}, \binits{J.}},
\bauthor{\bsnm{Zhang}, \binits{Y.}},
\bauthor{\bsnm{Hu}, \binits{Z.-D.}}:
\batitle{Universal decay of quantumness for photonic qubits carrying orbital
  angular momentum through atmospheric turbulence}.
\bjtitle{IEEE Photonics Journal}
\bvolume{10}(\bissue{1}),
\bfpage{1}--\blpage{9}
(\byear{2017})
\end{barticle}
\endbibitem
\bibitem{yang2022influence}
\begin{barticle}
\bauthor{\bsnm{Yang}, \binits{D.}},
\bauthor{\bsnm{Hu}, \binits{Z.-D.}},
\bauthor{\bsnm{Wang}, \binits{S.}},
\bauthor{\bsnm{Zhu}, \binits{Y.}}:
\batitle{Influence of random media on orbital angular momentum quantum states
  of optical vortex beams}.
\bjtitle{Physical Review A}
\bvolume{105}(\bissue{5}),
\bfpage{053513}
(\byear{2022})
\end{barticle}
\endbibitem

\bibitem{wang2004quantum}
\begin{barticle}
\bauthor{\bsnm{Wang}, \binits{X.-B.}}:
\batitle{Quantum error-rejection code with spontaneous parametric
  down-conversion}.
\bjtitle{Physical Review A}
\bvolume{69}(\bissue{2}),
\bfpage{022320}
(\byear{2004})
\end{barticle}
\endbibitem

\bibitem{knill1997theory}
\begin{barticle}
\bauthor{\bsnm{Knill}, \binits{E.}},
\bauthor{\bsnm{Laflamme}, \binits{R.}}:
\batitle{Theory of quantum error-correcting codes}.
\bjtitle{Physical Review A}
\bvolume{55}(\bissue{2}),
\bfpage{900}
(\byear{1997})
\end{barticle}
\endbibitem

\bibitem{suter2016colloquium}
\begin{barticle}
\bauthor{\bsnm{Suter}, \binits{D.}},
\bauthor{\bsnm{{\'A}lvarez}, \binits{G.A.}}:
\batitle{Colloquium: Protecting quantum information against environmental
  noise}.
\bjtitle{Reviews of Modern Physics}
\bvolume{88}(\bissue{4}),
\bfpage{041001}
(\byear{2016})
\end{barticle}
\endbibitem

\end{thebibliography}

\section*{Acknowledgement}
The authors thank the anonymous referee for valuable comments, which have helped us in improving the quality of the manuscript to a large extent.
Rajni thanks UGC for funding her research in the initial stages. Sooryansh thanks CSIR (Grant no.: 09/086 (2017)-EMR-I) for funding his research.
\section*{Data availability statement}
Data sharing not applicable to this article as no datasets were generated or analysed during the current study.
\section*{Disclosures}
The authors declare no conflicts of interest.
\section*{Author contribution statement}
R.B. and V.R. conceived the idea and contributed in the entire work. S. A. has contributed significantly in construction of quantum error correction code and in showing the equivalence of generalised flip channels and crosstalk channels. All the authors have written the manuscript collectively.

\end{document}